\documentstyle[aps,twocolumn]{revtex}
\input{epsf}
\renewcommand{\baselinestretch} {1.35}
\def\be{\begin{eqnarray}}
\def\ee{\end{eqnarray}}
\def\a{\alpha}
\def\b{\beta}
\def\d{\delta}
\def\h{\chi}
\def\g{\gamma}
\def\G{\Gamma}

\def\l{\lambda}
\def\o{\omega}
\def\ag{a^{\dagger}}
\def\bg{b^{\dagger}}
\def\cg{c^{\dagger}}

\def\pa{\partial}
\def\s{\sigma}
\def\ur{\uparrow}
\def\dr{\downarrow}
\def\bg{b^{\dagger}}
\def\cg{c^{\dagger}}
\def\e{\epsilon}

\begin{document}
\title{Models of spin structures  in $Sr_2RuO_4$.} 

\author{ A.A.Ovchinnikov$^{1,2}$, ~~M.Ya.Ovchinnikova$^1$}
\address{\it $^1$Joint Institute of Chemical Physics, Moscow.\\
$^2$Max Planck Institute for Physics of Complex Systems, Dresden.\\}


\wideabs{
\maketitle
\begin{abstract}
The mean field study of bands and the Fermi surfaces (FS) are carried out for
the 3-band model of $RuO_4$ plane of ruthenate for various spin structures. In
particular, the lowest spiral state with incommensurate vector
$Q=2\pi ({1\over 3},{1\over 3})$, corresponding to nesting of bands,
displays a set of FS's with electron pockets for $\g$ band and additional
shadow sheets along $\G -M$ lines for $\a (\b )$ bands. The latter may provide
a new interpretation of the dispersionless peaks (so called SS features)
observed in photoemission. For spiral state the polarisation asymmetry of
the photoemission intensity is revealed. The calculated spin
susceptibility for this state describes  the low frequency magnetic peak in
$\h''(q\o)$ at $q=Q$ observed in neutron scattering.

\end{abstract}
}


A single-layer quasi-2D ruthenate attracts attention as a superconductor 
($T_c\sim 1K$)  with a possible triplet type of pairing \cite{1,2}.
One of the arguments in favour of such type of pairing is the Knight shift
behaviour \cite{3}. It is suggested also that the pairing is governed by the
ferromagnet (FM) fluctuations 
which are assumed in an analogy with the ferromagnetism of the parent cubic
$SrRuO_3$.  A knowledge of the normal state electronic and magnetic properties
is a base for study the superconducting  pairing. The band structure and
magnetic fluctuations have been studied in \cite{4,5,6}. The substituted
compounds 
$Ca_{2-x}Sr_xRuO_4$ exhibit a complex phase diagram governed by the lattice
distortions. It includes the phases of the paramagnet (PM) or FM  metal or
antiferromagnet (AFM) dielectric. Contrary to such behaviour the lattice of the
$Sr_2RuO_4$ remains undistorted. The results on inelastic neutron
scattering (INS) show magnetic peak at incommensurate momentum
$Q\sim (0.3,0.3,1)$ (in units $({2\pi \over a},{2\pi \over a},
{2\pi \over c})$ \cite{7}). The static structure with the same period was
observed in substituted $Sr_2Ru_{1-x}Ti_xO_4$ \cite{9}.
In \cite{4,5} the peak position is connected with a structure of valence
bands, namely with a nesting of $\a$ and $\b$ bands of $d_{xz}$ and $d_{yz}$
nature, respectively. The nesting means an existence of the planar FS sections
which coincide with each other under a translation $k\to k+Q$ on a
nesting vector $Q$. 

There are many ways to describe the correlated systems with nesting. One
of them uses as a starting points the zero magnetic susceptibility
$\h_0(q,\o )$ of Lindhard \cite{10,11}. Nesting at some $q=Q$ leads to maximum
in $\h_0''(q,\o )$ at $\o\to 0$ and $q\to Q$. The subsequent renormalisation of
type $\h(q,\o)=\h_0/[1-J(q)\h_0]$  or $\h(q,\o)=\h_0/[1-U\h_0]$  in the t-J
or Hubbard model can explain the large observed density of the low
frequency excitations only if the denominators in renormalised
$\h$ is small at $q=Q$. This indicates on the possible  instability of the
system with respect to formation of the spin structure with given $q=Q$. Then
the renormalisation of $\h$ on base of zero spectrum may be inadequate.

Another way of treating the nesting in correlated system is to accept  a
specific spin structure, which removes the induced instability, and then 
to carry out the mean field (MF) consideration taking into account the assumed
spin order in zero approximation. In applications to cuprates it implies the
using of the AFM or spiral states as a zero approximation
\cite{12,13,14}. Usually the homogeneous MF solutions are considered which
supposes the infinite range of the spin order. In reality the doped compounds
have a finite range of the spin correlations. Therefore the homogeneous
MF solutions  must be thought rather as local or temporary structures in
system. A set of the possible spin structures for ruthenates is discussed.

Main features of bands and FS's for $Sr_2RuO_4$ obtained from the magnetic
quantum oscillations \cite{15} and from the photoemission (ARPES)
\cite{16,17,18,19} are reproduced by a three-band strong coupling model
proposed in \cite{4,5,6}. However, the ARPES data reveal a set of shadow FS's
which may indicate on a periodic spin structure in system. In particular, the
sharp dispersionless peak, called in \cite{19} the SS features, has been
observed and connected with a FS along lines $\G(0,0)-M(\pi,0)$. 
Interpretation of them  is controversial yet. In 
\cite{19} the SS features are attributed to the surface  states arising due
to a reconstruction of the surface layers. Such interpretation is argued by
the fact that the temperature cycling removes SS features and other shadow
FS's, i.e. destroys the surface superlattices.

The aim of present work is to study how the various spin structures manifest
in the form of the FS, in properties of shadow sheets of FS and in
magnetic susceptibility of 3-band model of $Sr_2RuO_4$. The calculations are
carried out in mean-field (MF) approximation. The PM, FM, AFM structures and
spiral states with incommensurate vector $Q=2\pi ({1\over 3},({1\over 3})$ are
considered. Such a spirality vector is chosen since it removes the nesting
instability of both $\a$ and $\b$  bands simultaneously. The possible bulk
origin of the dispesionless SS features along $\G -M$ line is discussed. They
are explained by Umklapp processes in periodic spiral structure. Calculations
of spin susceptibility $\h (q,\o)$ is done on base of spiral ground state as a
zero order state in RPA treatment. Thus, contrary to the standard
renormalisation of $\h$, the part of interaction is taken into account in a
non-perturbative manner. 

The valence bands in $Sr_2RuO_4$ is determined by electrons of $RuO_4$
plane. In approximate ionic model of $RuO_4=Ru^{4+}(d^4)(O^{2-})_4$ 
the four electrons occupy three lower d- orbits $d_{xy},~d_{xz},~d_{yz}$ 
of $t_{2g}$ symmetry.  These orbits together with $p_\pi$ orbits of oxygen
generate three valence bands with the total occupancy $n=4$ of electrons per
unit site of $RuO_4$ plane. The strong coupling Hamiltonian for corresponding
$\a,~\b$ and $\g$ bands of the $xz,~yz$ and $xy$ nature are  \cite{6}
\be
H=T+H_U;~~T=\sum_{\nu,\s}\sum_k \e_{\nu k}\cg_{\s\nu k}c_{\s\nu k}+T_{12}
\label{1}
\ee
$$
H_U=\sum_{n,\nu } \Bigl\{ U n_{\ur\nu n}n_{\dr\nu n} +
\sum_{\nu '\neq\nu}[U_2{1\over 4}n_{\nu n}n_{\nu' n}
-J{\bf S}_{\nu n}{\bf S}_{\nu' n}] \Bigr\}
$$
Here $U_2=2U-5J$ and 
$n_{\nu n}=n_{\nu\ur n}+n_{\nu\dr n}$, $n_{\nu\s n}$, ${\bf S}_{\nu n}$
are the site operators of the occupancy and the spin corresponding to the 
$\a,\b,\g$ bands ($\nu=1,2,3$). Without interband hopping the zero band
energies are 
\be
\e^0_{\nu k}=-t_0^{\nu}-2t^{\nu}_x \cos{k_x}-2t^{\nu}_y \cos{k_y}
+4t^{\nu}_{xy} \cos{k_y}\cos{k_y}
\label{2}
\ee
For the model parameters $t_0,t_x,t_y,t_{xy}$,$U$,$J$ in (1,2) we use the
values from \cite {6}. In undistorted lattice the interband coupling can be
only of the form
$T_{12}=\sum_{k,\s}4t_{\a\b}\sin{k_x}\sin{k_y}(\cg_{\s 1 k}c_{\s 2
k}+h.c.)$. A small value of $t_{\a\b}\sim t_{xy}^{1(2)}\sim 0.01 eV$ is
expected. 

We search the MF solutions of model (1) with definite order of the local spins
$<S_n>$. In absence of the spin-orbit interaction the orientation of
$<S_n>$ is arbitrary. For simplicity we consider that the spin
quantisation axis $z$ coincides with the crystal $c$ axis. With such agreement
we define the following set of the average spins in different spin structures
\be
<{\bf S}_n^{\nu}>=d_{\nu}[{\bf e}_x\cos{Qn}+{\bf e}_y\sin{Qn}]
\label{3}
\ee
For the considered SP, FM and AFM structures the corresponding vectors are 
\be
Q=2\pi({1\over3},{1\over3})~, ~~~Q=(0,0)~, ~~~
Q=(\pi,\pi)~
\label{4}
\ee
For our choice of the quantisation axis the MF equations have similar form for
all the states and they differ only in  $Q$. Repeat that we choose the vector
$Q$ of spiral state in such way that it removes of the nesting instability for
both quasi-1D $\a$ and $\b$ bands simultaneously.

The MF state  $\Phi_Q$ is determined by occupancy of the eigenstates  
$\bg_{\l k}$  (with spectrum $E_{\l k}$) of the linearised Hamiltonian 
$H_{lin}$
\be
[H_{lin},\bg_{\l k}]=E_{\l k}\bg_{\l k};~~~
\label{5}
\ee
In the renamed basis set $\ag_{i\nu k}$, $i=1,2$, the corresponding
eigen-operators are expressed as 
\be
\bg_{\l k}=\ag_{i\nu k}S_{i\nu,\l}(k) ; ~
~\ag_{i\nu k}=\{\cg_{\ur\nu k}, \cg_{\dr\nu k+Q} \}_i
\label{6}
\ee
Here $\l=1,\ldots,6$; the index $i=1,2$ corresponds to the spin projections
and the $\nu=1,2,3$ numerates the $\a,\b,\g$ bands.

The energy ${\bar H}({\bar T},r_{\nu},d_{\nu})$ averaged over the MF state
with the spin order (3) is a function of one-electron means of a density
$r_{\nu}$, of a local spin $d_{\nu}$ and of the mean kinetic energy ${\bar T}$
per site. In the basis set (6) the operators corresponding to the means 
$r_{\nu}$, $d_{\nu}$ are  
\be
\begin{array}{ll}
{\hat r}_{\nu}=&{1\over 2N}\sum_{i=1,2}\sum_k\ag_{i\nu,k}a_{i\nu,k}\\
{\hat d}_{\nu}=&{1\over 2N}\sum_{i=1,2}\sum_k\ag_{i\nu,k}a_{3-i,\nu,k}\\
\end{array}
\label{7}
\ee
Then the linearised Hamiltonian in (5) is equal to 
\be
{\hat H}_{lin}={\hat T}+{{\pa{\bar H}} \over{\pa r_{\nu}}}{\hat r}_{\nu}
+{{\pa{\bar H}} \over{\pa d_{\nu}}}{\hat d}_{\nu}+const
\label{8}
\ee
It determines the levels and eigen-operators in MF solution under a search. 

A study of the main and shadow FS for each band $\nu$ includes the calculation
of the one-electron spectral functions $A_{\nu}(k,\o)$ at $\o=0$. In MF
approximation one obtains
\be
A(k\o )=\sum_{\s}A_{\s}(k\o )
\label{9}
\ee
\be
\begin{array}{ll}
A_{\ur}(k\o )=&\sum_{\nu\l} |S_{1\nu ,\l}(k,\o)|^2
{\tilde \d}(E_{\l k}-\mu-\o)\\
A_{\dr}(k\o )=&\sum_{\nu\l} |S_{2\nu ,\l}(k-Q,\o)|^2
{\tilde \d}(E_{\l,k-Q}-\mu-\o)\\
\end{array}
\label{10}
\ee
Here $E_{\l} (k)$, $S_{i\nu,\l}(k)$ are defined by (5) and the $\d$-function 
is replaced by a broadened Lorentz-like function ${\tilde \d}(\e)$ with width
$\eta$ or by a function ${\tilde \d}(\e)=(4\eta)^{-1}\cosh^{-2}(\e/2\eta)$. 

Table below gives the values of energy (per site), the ratios of densities 
$2r_\a/r_\g$  and the values of local spins on the $\a,\b,\g$ orbits for the
spiral, FM, AFM and PM states of model with parameters taken from \cite{6} and
$t_{\a\b}=0$.  
$$
\begin{array}{l|llll}
           & SP       & FM     & AFM    & PM   \\
\hline
{\bar H},[eV]   & 2.708 & 2.7469 & 2.7552 & 2.7570\\
2r_\a/r_\g & 1.935 &  2.15   & 2.137  & 2.04  \\
d_\a(=d_\b)& 0.234 & 0.254  & 0.078  & 0     \\
d_\g       & 0.160 & 0.181  & 0.093  & 0     \\
\end{array}
$$
The states are listed in order of increasing of their energies. The lowest
state 
is a spiral one with vector $Q$, which removes the nesting instability for both
$\a$ and $\b$ bands. For all structures a value of the total local spin is of
order of $<S>=\sum_{\nu}d_{\nu}\sim 0.6$. It is consistent with that obtained
earlier for FM and AFM states. Note that all results refer to
undistorted lattice.

In spite of small difference in the energies the states have very different
structures of FS. For PM state one obtains the standard quasi-1D sheets of FS
at $k_{x(y)}=\pm 2\pi /3$  for $\a,\b$ bands and the electron-like FS for $\g$
band. The FS structure is in agreement with results of LDA band calculations
\cite{20,21} and with analysis of magnetic quantum oscillations.
Similar picture of FS has just been obtained \cite{19} from
photoemission (ARPES) data for the samples undergone the temperature
cycling in range $T=10-200 K$, which removes the shadow FS from the observed
map of FS. In \cite{19} it has been suggested that the cycling  destroys the
surface superstructures responsible for the shadow FS.

A set of FS's  for the FM and AFM states is presented in Fig.2 where the
images of spectral function $A(k_x,k_y,\o=0)$ at $\o=0$ are shown.
Corresponding broadening parameter $\eta$ in ${\tilde \d}$-function in (9)
is $\eta=0.02~eV$. For FM state one has a double set of FS since the bands
for the up and down spins are different.
In AFM state the FS of the $\g$ band are the boundaries of the electronic
pockets which are similar to the hole pockets of cuprates. Contrary to
cuprates, here the shadow edges of the pockets are are located on  inner sides
of the pockets closer to the centre of zone $\G(0,0)$.
Due to magnetic doubling of elementary unit there exist also the shadow images
of the main FS's reflected with respect to the lines $k_x\pm k_y=\pm\pi$. Note
that such reflected shadow FS's are really observed in ARPES map of
$Sr_2RuO_4$. But the same shadow FS may originate from several other possible
structures - from the surface (instead a bulk)  AFM order or from a lattice
reconstruction of the surface layers, as suggested  in \cite{19}.

Fig.~2 presents the FS's for the spiral state with $Q=2\pi (1/3,1/3)$. Several
specific features must be emphasised. 

1) The images of spectral functions (9)
corresponding to the up and down spin polarisations differ drastically from
each other. For instance, the 
image of $A_{\ur}(k_x,k_y,\o=0)$ displays only a half of all FS's seen in a
map of the total spectral function (see. Fig.2). For opposite polarisation one
has an  inverted structure of FS's since
$A_{\dr}(k_x,k_y,\o)=A_{\ur}(-k_x,-k_y,\o)$. Such polarisation asymmetry   
takes place only for the spin polarisations along the axis $z$ which is normal
to the plane of the local spin rotation in spiral state. This asymmetry in
the photoemission intensities might be observable by methods of the
polarisation photoemission only if an orientation of the spiral structure
relative to the lattice remains fixed by a spin-orbit interaction.

2) The shadow FS's appear approximately
along  the lines $k_x=0$ or $k_y=0$. They can be explained by the 
Umklapp processes in periodic spiral structure with $Q=2\pi (1/3,1/3)$ since
they coincide with the main $\a,~\b$ FS's  after a shift $q\to q\pm Q$. The
position of these shadow FS almost coincides with position of the
dispersionless peaks observed in photoemission along lines $\G(0,0)-M(\pi
,0)$\cite{19} and named the SS features. In \cite{19} the origin of these
features has been connected with electronic surface states. It was supposed
that these states arise due to a lattice reconstruction of the
surface layers  $RuO_4$ with doubling  of elementary unit.
Similar lattice distortion with staggered rotation
of octahedra was observed in FM phase of the substituted compounds. One of
arguments in favour of surface origin of SS patterns was a disappearance of 
them and the other shadow FS's after the temperature cycling of samples. Our
calculations of spiral states allows to suggest a bulk origin of SS features
due to spiral spin  structure. Then disappearance of the SS features after the
temperature cycling might  mean a destroying the spiral spin (instead of
surface) order in system. 

It has been found also the MF solution with the periodic spin and charge
structure containing 18 lattice sites in the elementary unit. Its MF energy is
slightly higher than that of spiral state, but lower than energies of FM, AFM,
PM states of model. In this solution the FS's of $\a,\b$ bands disappear due
to formation of a gap along all former FS's of this bands and the
FS of $\g$ band has very complicated form. Both pecularuties are inconsistent
with the observed FS's for all three bands.

It was interesting also to calculate the spin susceptibility $\h(q,\o )$ of
the three-band system in spiral state. For this aim we apply the RPA approach
taken in a form different from the standard RPA. Our method uses the MF state
with the spiral spin structure as a zero approximation. Thus the nesting
instability is removed nonperturbatively already in zero order
approximation. The consideration is inspired by similar treatment of the
one-band  
Hubbard model \cite{12,13,14}. The details of calculations will be given
elsewhere. Here in Fig.~3 we show  only the calculated $Im\h(q,\o)$ at small
$\o=0.02~eV$ for q varying along the contour $\G (o,o)-M(\pi,0)-Y_1(\pi,\pi
)-\G-Y_2(-\pi,\pi)$ (see insert in Fig.3). Besides the expected absorption peak
at $q=Q$, there is almost symmetric peak near the $q=2\pi (-1/3,1/3)$ . Both
peak positions are in accordance with the positions of the observed magnetic
peaks in INS on $Sr_2RuO_4$ at $q\sim2\pi(1/3,1/3,1)$. Note that an integral
intensity ${\tilde \h}(\o )=\int\h ''(q\o )d^2q/4\pi^2$ occurs to be of order
$\sim 1~eV^{-1}$ which is comparable with the similar quantity  
$\sim 2~eV^{-1}$ in cuprates \cite{22}.

In conclusion, the main and shadow sheets of FS are obtained for the PM, FM,
AFM and spiral states of three-band model of $RuO_4$ plane of
$Sr_2RuO_4$. The map of FS's is highly specific to the spin
structure. The lowest MF state appears to be the spiral state with the
spirality vector $Q=2\pi({1\over 3},{1\over 3})$ removing the nesting
instability for both quasi 1D ($\a$ and $\b$) bands simultaneously.
For spiral state the main FS's display in photoemission with intensity
depending on the spin polarisation. Such polarisation asymmetry of
photoemission spectra may be a test for the spiral structure in
$Sr_2RuO_4$. In spiral state the shadow FS's along $\G -M$ line are
revealed. Coincidence of their position with the position of the
dispersionless SS peaks in ARPES map of $Sr_2RuO_4$ allows to suggest the bulk
origin of the observed SS features as caused by Umklapp processes in spiral
state. The calculation of the spin susceptibility $\h(q,\o)$ for spiral state
confirms the existence of the magnetic peak in the low frequency spin
excitations at $q\sim Q$  in accordance with the INS observations. 
If the spiral structure does exist in ground state of $Sr_2RuO_4$ the
new questions can be put forward on the symmetry of superconducting
pairing, on phase of quantum magnetic oscillations and so on. 

Work is supported by Russian Fund of Fundamental Research (Projects
No. 00-03-32981 and No. 00-15-97334). Authors thanks V.Ya Krivnov for useful
discussions and P.Fulde for possibility to work in
Max Planck Institute for Physics of Complex Systems, Dresden.


\renewcommand{\baselinestretch} {1.00}

\vspace {0.1in}
{\bf Captions to Figures}
\vspace {0.05in}

Fig.~1.
The Fermi surfaces for the  FM (left) and AFM (right) spin structures of
three-band model obtained as image of spectral function (9) at $\o=0$ and the
broadening parameter $\eta=0.02eV$. The model parameters are those from
\cite{6} and $t_{\a\b}=0.01 eV$.  

Fig.~2.
The Fermi surfaces of  system in spiral state with vector 
$Q=2\pi({1\over3},{1\over 3})$:  1) the image of spectral function 
$A_{\ur}(k_x,k_y,\o=0 )$ for the spin-up polarisation (left), 2) the same for
the total spectral  function (9) (right). The parameters are those from
\cite{6} and $t_{\a\b}=0$ . Some nonlinear transformation of $A(k,\o$ is
applied to strength the shadow bands. 

Fig.~3.
The imaginary part of spin susceptibility $\h''(q,\o)$ averaged over
polarisations (full curve) and its components  $\h''_{xx}(q,\o)$ (dashed)
and $\h''_{zz}(q,\o)$ (dotted) at $q$ varying along the contour
$\G-Y_1 -M-G-Y_2$ (see insert). The  frequency and the broadening parameter
are $\o=\eta= 0.02eV$.

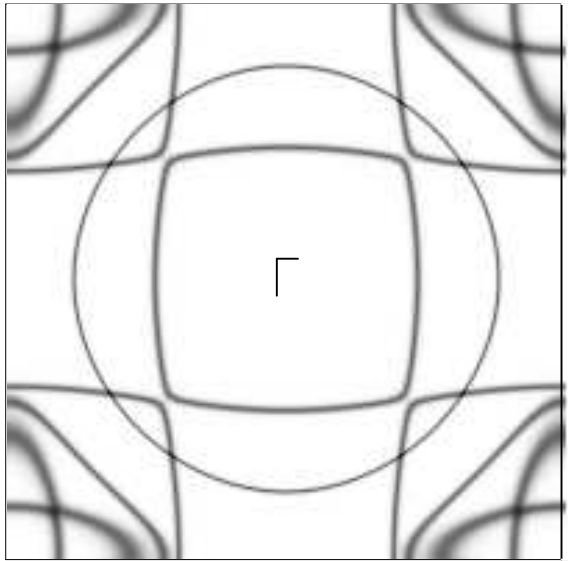
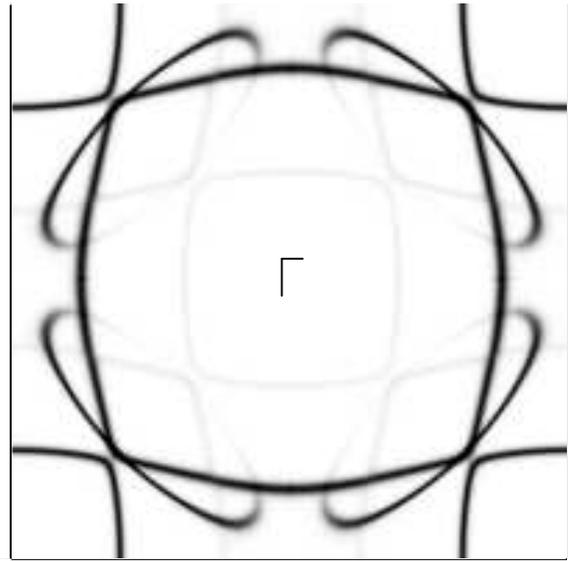
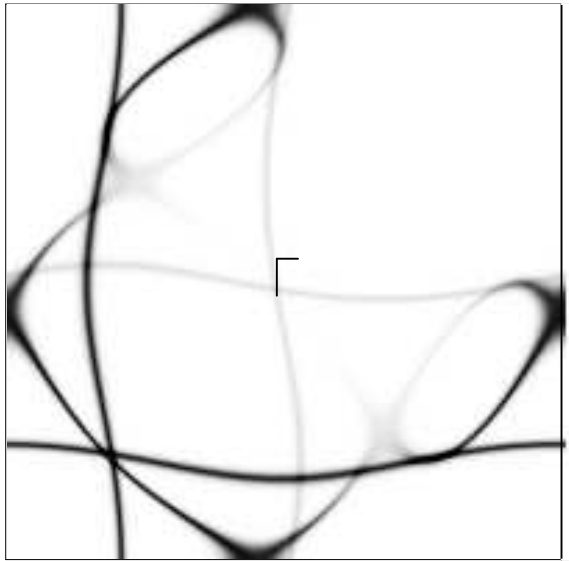
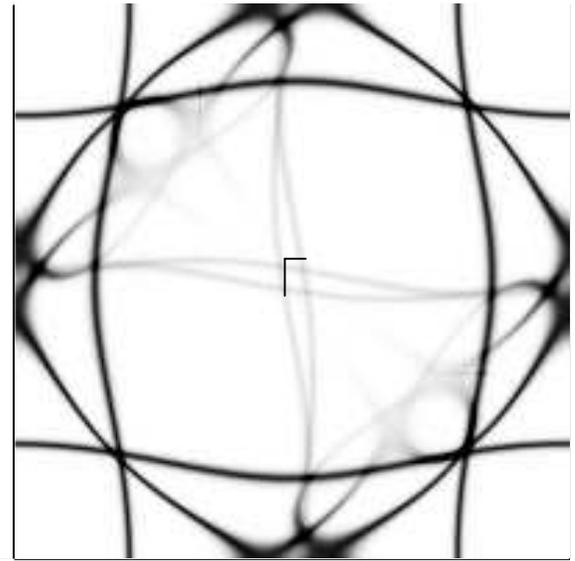
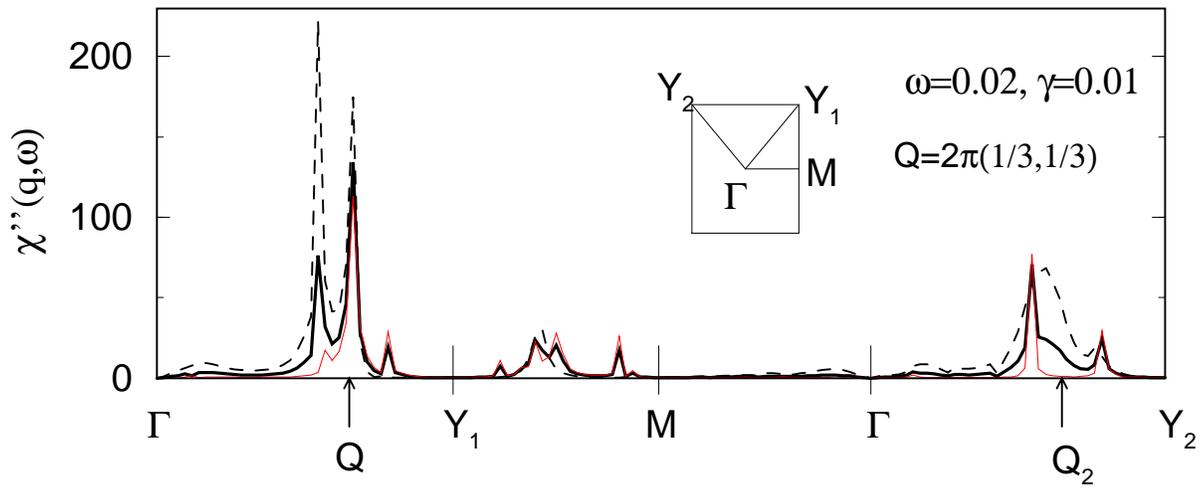
\end{document}